\documentclass[graybox]{article}

\usepackage{mathptmx}       % selects Times Roman as basic font
\usepackage{helvet}         % selects Helvetica as sans-serif font
\usepackage{courier}        % selects Courier as typewriter font
\usepackage{type1cm}        % activate if the above 3 fonts are

\usepackage{amsmath}

\usepackage{graphicx}
\usepackage{epsfig}
\usepackage{epstopdf}         % don't forget the -shell-escape for pdftolatex
\usepackage{subfigure}
\usepackage{multirow}
\usepackage{xcolor}

\usepackage{filecontents}

\usepackage{url}
\usepackage{verbatim}
\usepackage{fancyvrb}
\fvset{fontsize=\footnotesize}
\RecustomVerbatimEnvironment{verbatim}{Verbatim}{}

\usepackage{pgfplots}
\usepackage{color}

\usepackage[section]{placeins} %  prevents placing floats before a section

\usepackage[OT2, T1]{fontenc}
\usepackage[russian, english]{babel}

\DeclareRobustCommand{\rchi}{{\mathpalette\irchi\relax}}
\newcommand{\irchi}[2]{\raisebox{\depth}{$#1\chi$}} % inner command, used by \rchi

\usepackage{subfigure}
\usepackage{subfig}
\usepackage{authblk} % para autores con diferente afiliacion

\newcommand*{\stext}[1]{\text{ #1 }}

\begin{document}

% box of limited width for the verbatim
\newsavebox\myv

\title{An oracle-based attack on \\ CAPTCHAs protected against \\ oracle attacks.}

\author[1]{Carlos Javier Hern\'andez-Castro\thanks{chernandez@ucm.es}}
%\author[4]{Julio C\'esar Hern\'andez-Castro\thanks{jch27@kent.ac.uk}}
\author[2]{Mar\'ia D. R-Moreno\thanks{mdolores@aut.uah.es}}
\author[3]{David F. Barrero\thanks{david@aut.uah.es}}
\author[4]{Shujun Li\thanks{shujun.li@surrey.ac.uk}}
\affil[1]{Universidad Complutense, Madrid, Spain}
%\affil[4]{School of Computing, University of Kent, Canterbury, U.K.}
\affil[2]{Universidad de Alcal\'a, Madrid, Spain}
\affil[3]{Universidad de Alcal\'a, Madrid, Spain}
\affil[4]{University of Surrey, Guildford, UK}

\renewcommand\Authands{ and }

\maketitle
					                                
\begin{abstract}
CAPTCHAs/HIPs are security mechanisms that try to prevent automatic abuse of services. They are susceptible to learning attacks in which attackers can use them as oracles. Kwon and Cha presented recently a novel algorithm that intends to avoid such learning attacks and ``detect all bots''  \cite{Kwon16}. They add uncertainties to the grading of challenges, and also use \textit{trap images} designed to detect bots. The authors suggest that a major IT corporation is studying their proposal for mainstream implementation. 
We present here two fundamental design flaws regarding their \textit{trap images} and \textit{uncertainty grading}. These leak information regarding the correct grading of images. Exploiting them, an attacker can use an UTS-CAPTCHA as an oracle, and perform a learning attack. Our testing has shown that we can increase any reasonable initial success rate up to $100\%$.
\end{abstract}

%\keywords{CAPTCHA, HIP, Uncertainty}

\section{Introduction}
\label{sec:introduction}

Free on-line services have become prevalent since the broad deployment of the Internet in the late 90's. The abuse of such services, using automated methods, can be the first step towards more sophisticated attacks that can result in significant revenue for the attackers. Naor \cite{Naor96} was the first to propose a theoretical security framework based on the idea of discerning humans from bots using problems that could be solved easily by humans but were thought to be hard for computers. 

Many CAPTCHAs, including the currently most used one, are based on image classification. They require the user to tell which images from a set pertain to a particular category (\cite{Shet14incrc}). Many of these proposals have been analyzed and broken (\cite{Golle09, Homakov14, Sivakorn16blackhat}). 

Most types of the CAPTCHAs based on classification, including image classification CAPTCHAs, are susceptible to a learning attack. To perform one, all we need is a bot based on a random or a very weak classifier that can successfully pass just a few challenges, even if mostly by chance. Once a challenge is successfully solved, the bot can learn the correct classification of the images in that challenge. Thus, the bot can use the CAPTCHA as an oracle and dramatically increase its success rate. %(citar art).

In \cite{Kwon16}, Kwon and Cha proposed a way to prevent bots from learning new examples using the CAPTCHA service as an oracle. To do so, they employ two mechanisms. First, they add uncertainty to a grading function: when they grade an answer to a challenge, they do not take all images in the challenge into consideration. Second, they use what they call ``trap images'' to detect ``all bots''. Using these two mechanisms, they propose to strengthen any image-based CAPTCHA into what we will call an \textit{Uncertainty \& Trap Strengthened-CAPTCHA} (UTS-CAPTCHA).

These protection mechanisms can play a fundamental role in enabling image CAPTCHAs to be secure. This additional protection is of particular relevance because many Deep Learning architectures, like DCNNs, are getting extremely successful at classifying images \cite{Krizhevsky12,Ciregan12multi,Goodfellow13mdnrfsviudcnn,Taigman14,Stark15captcha,Sivakorn16}. 

DL approaches have some limitations \cite{Nguyen15,Papernot15,Osadchy16}, but as of now, DL classifiers remain a serious threat to image classification CAPTCHAs.

In this article, we present an attack against the scheme introduced by Kwon and Cha \cite{Kwon16}.

\section{UTS-CAPTCHA}     
\label{sec:UACAPTCHA}

Kwon and Cha \cite{Kwon16} propose a method to build a UTS-CAPTCHA from a typical image classification CAPTCHA. In their example, they start from a one-class image classification-based CAPTCHA. This CAPTCHA presents $22$ images per challenge, containing faces of people. Some of them are pictures of a popular person (\textit{Bill Gates}, co-founder of Microsoft Corp.). They divide their face images into two groups: $M$ for those depicting \textit{Bill Gates} ($M$ for ``M''ust be picked) and $NM$ for those depicting other people ($MN$ for ``M''ust ``N''ot be picked). 

Then they \textit{strengthen} it using two mechanisms:
\begin{itemize}
\item For each challenge, they select a random number of images between $0$ and $8$ that are not going to be graded ($NE$). The challenge can be passed if the user answers correctly to the other images (in $C - NE$). Thus the users' classification of images in $NE$ will be irrelevant towards marking the challenge as passed or not.
\item If a user $u$ passes a challenge answering incorrectly to some images, these are included in a set $TI_u$ of \textit{trap images} for that user. Following challenges will include $1$ or $2$ of these \textit{trap images} in $TI_u$, and they will always be used for grading.
\end{itemize}

In general, we can model a typical image classification-based CAPTCHA that uses only one class  as $iCAPTCHA = (M, MN, |C|)$, where $M$ is the set of images that pertain to the class, $MN$ are other images that are not from the class, and $|C|$ is the number of images used per challenge $C$. 

This is a simplified model, as a general image-based CAPTCHA can be multi-class, |C| does not need to be fixed, etc. This limitations do not affect the method proposed to turn an image-based CAPTCHA into an UTS-CAPTCHA, nor they affect our proposed attack.

The idea behind their proposal is that once a bot is miss-classifies an image in $NE$, and it is included in $TI$, the bot will continue to miss-classify it consistently.

Kwon and Cha add an image $x_i$ to the trap images set $TI$ only when the challenge $C$ is correctly graded, but the image $x_i$ is incorrectly classified. This is possible as the user will still pass the challenge, if that image is among the randomly selected ones that will be irrelevant towards marking the challenge ($x_i \in NE_C$). 
As long as $TI \neq \emptyset$, Kwon and Cha always use the images in it. In each challenge there is one or two trap images. If present, trap images always count towards marking the challenge as passed or not. 

\section{Learning from trap images}     
\label{sec:LearningFromTrapImages}

It seems obvious that in Kwon and Cha's scheme, and as long as the intended purpose of detecting bots works, the size of $TI$ will be much smaller than the size of the rest of images on the database $|M \cup MN|$. This is so because ideally, following \cite{Kwon16}, $TI$ will be filled once and only once per each bot.

Note that this scenario presents a problem. For each challenge $C$, less images are taken from $TI$ ($1$ to $2$) than from $M \cup MN$ ($22$). Still, the size ratios of these two sets are much more different in both cases. That is: ${{1,5} \over {|TI|}} \neq {{|C|} \over {|M \cup MN|}}$. This ratio difference implies that for the challenges following the creation of $TI$, the images from $TI$ will repeat significantly more than the rest of images that appear in any challenge.

We can add a simple heuristic to a bot so it can detect trap images. It can be as easy as counting the appearances of images since their last apparition in a challenge that was passed. Then:
\begin{itemize}
\item if an image $x$ keeps appearing much more than expected if it was from $M \cup MN$ ($|C|/(|M|+|MN|)$), and
\item it has appeared in a challenge that was successfully solved, and 
\item in every challenge it has appeared since the above occasion, the challenge was not passed
\end{itemize}

Then this image is probably a trap image, $x \in TI$. Great news, as once we learn that an image is in $TI$, we know without a doubt the correct classification for it: the opposite to what we have been answering to it so far.

More precisely, we can run a Pearson's $\rchi^2$ test with $1$ degree of freedom where the target distribution is a Binomial in which the chances of $x_i$ appearing in a challenge are are ${|C|} \over {(|M|+|MN|)}$. When the $p-value$ of the test drops below a threshold (i.e. 0.05) we can label an image as a trap image.

We can estimate the probability of $Pr(x \in TI)$ based on its frequency of appearances using Bayes' theorem (Equation \ref{eq:BayesTheorem}).

\begin{equation} \label{eq:BayesTheorem}
\begin{split} 
Pr(x_1 \in TI|X=a(x_1,h_1,h)) = \\ 
{{Pr(X=a(x_1,h_1,h)|x_1 \in TI)*Pr(x_1 \in TI)} \over {Pr(X=a(x_1,h_1,h)|x_1 \in TI)*Pr(x_1 \in TI) + Pr(X=a(x_1,h_1,h)|x_1 \not\in TI)*Pr(x_1 \not\in TI)}}
\end{split}
\end{equation}

Where:

\begin{itemize}
\item $h_1$ : as we index challenges from $1$ onwards as they appear, $h_1$ is the highest challenge number at which we saw the image $x_1$, and that was passed by the bot. Since them, the bot has not passed any challenge that included $x_1$ on it (else, $x_1$ would not be in $TI$).
\item $h$ : total number of challenges until this moment.
\item $a(x_1,h_1,h)$ : number of appearances of $x_1$ in the  challenges from $h_1$ to $h$ ($h_1$ not counting).
\end{itemize}

Let's first solve each part of Equation \ref{eq:BayesTheorem}. $Pr(X = a(x_1,h_1,h)|x_1 \in TI)$ can be written as $Pr(x_1 \stext{has appeared} a(x_1,h_1,h) \stext{times in} h-h_1 \stext{challenges, given that} x_1 \in TI)$. For one challenge, the chances of $x_1$ appearing (vs. not appearing) in the challenge, if $x_1 \in TI$, are:

\begin{equation} \label{eq:PrXGivenA_2}
\begin{split}
Pr_{x1_{TI}} = \\
Pr(x_1 \stext{in a challenge} | x_1 \in TI) = \\
1 - {{{TIs-1} \choose eTIpc} \over {TIs \choose eTIpc}} = {eTIpc \over TIs} \approx {E(eTIpc) \over eTIs}
\end{split}
\end{equation}
%\stext{Unless} eTIs <= eTIpc, \stext{in that case it is = 1}.

As long as $0 <= E(eTIpc) <= eTIs-1$, if not, $Pr_{x1_{TI}} = 1$.

Where:

\begin{itemize}
\item $eTIpc$ : elements from $TI$ presented per challenge. $E(eTIpc)$ is the expected value, if $eTIpc$ is variable. In \cite{Kwon16}, they define $eTIpc_{min} = 1$ and $eTIpc_{max} = 2$, and randomly choose $eTIpc$ from that interval as long as the size of $TI$ allows for it.
\item $TIs$ : size of $TI$.
\item $eTIs$ : (attackers') estimation of the size of $TI$.
\end{itemize}

There are several approximations to $eTIs$, the estimated size of $TI$. We can limit it by considering the number of correctly solved challenges and the value of $\alpha$. Yet there are better approximations to it, some of which we are testing now.

We can similarly calculate $Pr(X = a(x_1,h_1,h)|x_1 \in TI)$ using the Binomial distribution, $Pr(x_1 \in TI)$ , $Pr(X=a(x_1,h_1,h)|x_1 \not\in TI)$, $Pr(X=a(x_1,h_1,h)|x_1 \not\in TI)$ also using the Binomial distribution and $Pr(x_1 \notin TI)$.

Kwon and Cha also propose that if an image from $TI$ is solved correctly (for a non-specified cause), it should be removed from $TI$, as it no longer serves to differentiate this bot. This removal mechanism further limits $|TI|$, thus increasing the repetitions of appearances of images from $TI$. Our attack benefits from this.

%We can use a simple procedure to calculate the upper bound of the size of $TI$ in every moment.

Given any bot $b$ that successfully solves the CAPTCHA with a particular rate $\beta$, Kwon and Cha comment that the upgraded version of the CAPTCHA, that we will call UTS-CAPTCHA will lower down the success rate of $b$ to $0$ once they start to present $b$ with trap images. 

We propose now an attack that can be based on any bot $b$ able to solve an image CAPTCHA with a minimal success rate ($\geq 1\%$). Our attack significantly improves $b$ success rate with the strengthened version UTS-CAPTCHA by not just avoiding trap images but in fact learning from then. 

\section{Attack design}     
\label{sec:AttackDesign}

Our attack leverages any given bot $b$ that successfully solves the original image CAPTCHA with a minimum success rate. Let $\beta$ be its particular success rate \footnote{This means that $b$ has an accuracy, as a classifier, of $\beta^{1 \over {|C|}}$.}. We will modify $b$, allowing it solve the uncertainty \& trap-augmented version of the original image-based CAPTCHA (UTS-CAPTCHA for short). We do so by creating $l_b$, an improved \textit{learning} bot based on $b$. $l_b$ can identify and learn from the trap images of UTS-CAPTCHA.

To do so, for each image $x_i$ that appears in a challenge $c_{h_i} = {x_1, ... x_{|C|}}$ we calculate the probability that $pr_{TI}(x_i) = Pr(x_i \in TI)$. 

To calculate it, we have to keep track of some values for each image $x_i$: the number of the last challenge that contained $x_i$ and that was correctly solved ($h_{(x_i,1)}$), the classification we gave for $x_i$ in $h_{(x_i,1)}$ ($ans_{h_1}(x_1)$) and the number of appearances of $x_i$ since challenge $h_{(x_i,1)}$ ($a(x_i,h_{(x_i,1)},h)$). We also have to register the number of the current challenge ($h$) and have an estimation of the size of $TI$ at challenge number $h$ ($eTIs_{h}$). This estimation can be done in several ways, for example using $|(x_i | pr_{TI}(x_i) \geq Th_{TI})|$.

With every new challenge $c_{h_i} = {x_{i_1}, ... x_{i_|C|}}$ received, we have to consider the increment in the number of times an image $x_i$ has appeared, for those images $x_i \in c_{h_i}$. We have to recompute $pr_{TI}(x_i)$ for every image in $c_{h_i}$. But as $pr_{TI}(x_i)$ only increases for the images $x_i \in c_{h_i}$, we might be able to restrict the updated computation to this set of images in $c_{h_i}$, unless we are using the whole set of $pr_{TI}(x_i)$ for something else, like estimating $|TI|$.

We set a \textit{knowledge-threshold} $KTh_{TI}$, above which we will consider that we have detected a new trap image \footnote{Note that there are more precise alternatives, as running a Pearson's $\rchi^2$ test with $1$ degree of freedom where the target distribution is a Binomial in which the chances of $x_i$ appearing in a challenge are are ${|C|} \over {(|M|+|MN|)}$}. Whenever $pr_{TI}(x_i) \geq KTh_{TI}$, we introduce this correction into our knowledge-base: $lc(x_i) = 1 - ans_{h_1}(x_1)$, where \textit{lc} is the ``learned classification'' %and $learned_probl_{TI}(x_i) = pr_{TI}(x_i)$
. $KTh_{TI}$ is our ``knowledge-threshold'' with a sensible value, like $0.9$. %No: esperamos a pasar una challenge para eso When we add $x_i$ to $lc$, we also do $h_{i_{x_i}} = h$ and reset $a(x_i,h_{i_{x_i}},h)$.

Now we create a bot $l_b$ that classifies images $x_i$ as $b$ does when $x_i \not\in lc$, but uses the knowledge gained to \textit{avoid} trap images. To do so, our bot classifies as per Equation \ref{eq:l_b_out}.

\begin{equation} \label{eq:l_b_out}
l_b(x_i) = \stext{if} x_i \in lc \stext{then} lc(x_i) \stext{else} b(x_i)
\end{equation}

Note that $x_i$ will remain in $TI$ until a challenge that includes $x_i$ is correctly solved by $l_b$. That will happen, as once we figure out that $x_i \in TI$, we start classifying $x_i$ as $1-b(x_i)$ and wait for it to appear into a challenge that is successfully solved by $l_b$. Only then we assume that $x_i \notin TI$.

\section{Attack results}     
\label{sec:AttackResults}

Kwon and Cha have implemented their UTS-CAPTCHA on-line. For each challenge, downloading of the images takes over a few seconds, so a learning test involving thousands of trials would take too much time to complete. We decided to perform all the testing on our simulation of their UTS-CAPTCHA.

\begin{figure}[!htb]
\centering
\includegraphics[scale=.25]{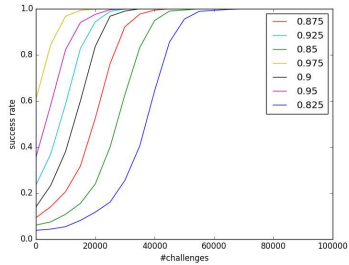}
\caption[Attack success rate per successive series of $5000$ experiments with $mr=5$.]{Attack success rate per successive series of $5000$ experiments when $mr=5$. }
\label{fig:SRPerBetaPrima}
\end{figure}

We are currently testing our attack with very promising results. Among our provisional results, Figure \ref{fig:SRPerBetaPrima} shows how the accuracy increases per each series of $5000$ experiments as we keep learning more trap images. As can be seen, leveraging a bot $b$ with a minimal success ratio $\beta=1.4\%$, we are able to achieve a success rate of $l_b=100\%$ with only $70K$ challenges.

\section{Learning with uncertainty}     
\label{sec:LearningWithUncertainty}

Kwon and Cha \cite{Kwon16} propose to add uncertainty to the evaluation of the challenges. In their design, this measure has two potential benefits: prevent bots from learning the correct classifications, even when a challenge has been passed, and allow for the introduction of \textit{trap images}. This measure also has one important effect: it reduces the strength of the CAPTCHA, as not all challenge sub-answers are graded.

This measure does not prevent a bot from successfully performing a learning attack, as the chances of passing a challenge are still correlated with the correct image classification. Full details of a similar attack can be built through the analysis of the correlations of image classifications and challenges solved.

\section{Conclusion}     
\label{sec:Conclusion}

Kwon and Cha \cite{Kwon16} present a way to increase the robustness of an image classification CAPTCHA, creating a \textit{Uncertainty \& Trap-Strengthened}-CAPTCHA (UTS-CAPTCHA). They propose to do so by preventing learning attacks and detecting bots. This is of particular relevance given the recent achievements in ML (\cite{Krizhevsky12,Ciregan12multi,Goodfellow13mdnrfsviudcnn,Taigman14,Stark15captcha,Sivakorn16}).

We show here that their design is flawed, as it still allows to perform a learning attack. We are currently testing our attack with very promising results, leveraging a bot with a minimal success ratio of $1.4\%$ up to $100\%$. 

We are currently analyzing whether there are protections for UTS-CATPCHAs from our attack or similar attacks, as well as additional attacks to UTS-CAPTCHAs.

%%\begin{acknowledgement}
%%\acks

%\section*{Acknowledgements}

%The first author wants to thank, in no particular order, Julio H.O., Declan J. H. O., Julio C. H.C., \foreignlanguage{russian}{Zhenya S.}, Carmen C. V. and \foreignlanguage{russian}{Lena S.} for their contributions and support. 

%\end{acknowledgement}

%\newpage
\bibliographystyle{plain}
\bibliography{main-arxiv}

\begin{thebibliography}{10}

\bibitem{Ciregan12multi}
Dan Ciregan, Ueli Meier, and J{\"u}rgen Schmidhuber.
\newblock Multi-column deep neural networks for image classification.
\newblock In {\em Computer Vision and Pattern Recognition (CVPR), 2012 IEEE
  Conference on}, pages 3642--3649. IEEE, 2012.

\bibitem{Golle09}
Philippe Golle.
\newblock Machine learning attacks against the asirra captcha.
\newblock In {\em Proceedings of the 5th Symposium on Usable Privacy and
  Security, SOUPS 2009, Mountain View, California, USA, July 15-17, 2009}, ACM
  International Conference Proceeding Series. ACM, 2009.

\bibitem{Goodfellow13mdnrfsviudcnn}
Ian~J. Goodfellow, Yaroslav Bulatov, Julian Ibarz, Sacha Arnoud, and Vinay~D.
  Shet.
\newblock Multi-digit number recognition from street view imagery using deep
  convolutional neural networks.
\newblock {\em CoRR}, abs/1312.6082, 2013.

\bibitem{Homakov14}
Egor Homakov.
\newblock The no captcha problem.
\newblock
  \url{http://homakov.blogspot.com.es/2014/12/the-no-captcha-problem.html},
  2014.

\bibitem{Krizhevsky12}
Alex Krizhevsky, Ilya Sutskever, and Geoffrey~E. Hinton.
\newblock Imagenet classification with deep convolutional neural networks.
\newblock In F.~Pereira, C.~J.~C. Burges, L.~Bottou, and K.~Q. Weinberger,
  editors, {\em Advances in Neural Information Processing Systems 25}, pages
  1097--1105. Curran Associates, Inc., 2012.

\bibitem{Kwon16}
S.~Kwon and S.~Cha.
\newblock A paradigm shift for the captcha race: Adding uncertainty to the
  process.
\newblock {\em IEEE Software}, 33(6):80--85, Nov 2016.

\bibitem{Naor96}
Moni Naor.
\newblock Verification of a human in the loop or identification via the turing
  test.
\newblock \url{http://www.wisdom.weizmann.ac.il/~naor/PAPERS/human.ps}, 1996.

\bibitem{Nguyen15}
Anh Nguyen, Jason Yosinski, and Jeff Clune.
\newblock Deep neural networks are easily fooled: High confidence predictions
  for unrecognizable images.
\newblock In {\em The IEEE Conference on Computer Vision and Pattern
  Recognition (CVPR)}, June 2015.

\bibitem{Osadchy16}
Margarita Osadchy, Julio Hernandez-Castro, Stuart Gibson, Orr Dunkelman, and
  Daniel P{\'e}rez-Cabo.
\newblock No bot expects the deepcaptcha! introducing immutable adversarial
  examples with applications to captcha.
\newblock {\em IACR Cryptology ePrint Archive}, 2016:336, 2016.

\bibitem{Papernot15}
Nicolas Papernot, Patrick McDaniel, Somesh Jha, Matt Fredrikson, Z~Berkay
  Celik, and Ananthram Swami.
\newblock The limitations of deep learning in adversarial settings.
\newblock In {\em Proceedings of the 1st IEEE European Symposium on Security
  and Privacy}, 2015.

\bibitem{Shet14incrc}
Vinay Shet.
\newblock Are you a robot? introducing no captcha recaptcha.
\newblock
  \url{https://security.googleblog.com/2014/12/are-you-robot-introducing-no-captcha.html},
  2014.

\bibitem{Sivakorn16}
Suphannee Sivakorn, Iasonas Polakis, and Angelos~D Keromytis.
\newblock I am robot:(deep) learning to break semantic image captchas.
\newblock In {\em 2016 IEEE European Symposium on Security and Privacy
  (EuroS\&P)}, pages 388--403. IEEE, 2016.

\bibitem{Sivakorn16blackhat}
Suphannee Sivakorn, Jason Polakis, and Angelos~D Keromytis.
\newblock {I'm not a human: Breaking the Google reCAPTCHA}.
\newblock {\em Black Hat}, (i):1--12, 2016.

\bibitem{Stark15captcha}
Fabian Stark, Caner Haz{\i}rbas, Rudolph Triebel, and Daniel Cremers.
\newblock Captcha recognition with active deep learning.
\newblock In {\em Workshop New Challenges in Neural Computation 2015}, page~94.
  Citeseer, 2015.

\bibitem{Taigman14}
Yaniv Taigman, Ming Yang, Marc'Aurelio Ranzato, and Lior Wolf.
\newblock Deepface: Closing the gap to human-level performance in face
  verification.
\newblock In {\em The IEEE Conference on Computer Vision and Pattern
  Recognition (CVPR)}, June 2014.

\end{thebibliography}
%\bibliographystyle{alpha}
%\bibliography{references}

% Activate the appendix
% from now on sections are numerated with capital letters

\end{document}